\documentclass{ws-p9-75x6-50}

\begin{document}

\title{Magnetic properties\\
of the doped two-dimensional antiferromagnet}

\author{A.~Sherman}
\address{Institute of Physics, Riia 142, 51014 Tartu, Estonia}

\author{M.~Schreiber}
\address{Institut f\"ur Physik, Technische Universit\"at, D-09107
Chemnitz \\
and School of Engineering and Science, International University
Bremen\\
Campus Ring 1, D-28759 Bremen, Federal Republic of Germany}

\maketitle

\abstracts{ The variety of the normal-state magnetic properties of
cuprate high-$T_c$ superconductors is interpreted based on the
self-consistent solution of the self-energy equations for the
two-dimensional $t$-$J$ model. The observed variations of the spin
correlation length with the hole concentration $x$, of the spin
susceptibility with $x$ and temperature $T$ and the scaling of the
static uniform susceptibility are well reproduced by the calculated
results. The nonmonotonic temperature dependence of the Cu spin-lattice
relaxation rate is connected with two competing tendencies in the
low-frequency susceptibility: its temperature decrease due to the
increasing spin gap and the growth of the susceptibility in this
frequency region with the temperature broadening of the maximum in the
susceptibility.}

Magnetic properties of cuprate perovskites have been extensively
studied during the last years, both because of their unusual behavior
and in the hope that they might provide insight into the physical
origin of high-$T_c$ superconductivity.\cite{Johnston}$^-$\cite{Fong}
These properties provide a good test for models proposed for the
description of cuprates and approximations used for their solution. One
of the most popular models is the two-dimensional $t$-$J$ model of Cu-O
planes. This model describes the low-energy dynamics of the realistic
three-band Hubbard model.\cite{Sherman98} Recently we have suggested an
approximate method for the solution of the $t$-$J$ model, which has
merits of retaining the rotation symmetry of spin components in the
paramagnetic state and of the absence of any predefined magnetic
ordering.\cite{Sherman02} The obtained magnetic properties of the model
appear to be close to those observed in the cuprate perovskites which
allows us to clarify mechanisms leading to the unusual behavior.

The Hamiltonian of the two-dimensional $t$-$J$ model reads
\begin{equation}
H=\sum_{\bf nm\sigma}t_{\bf nm}a^\dagger_{\bf n\sigma}a_{\bf
m\sigma}+\frac{1}{2}\sum_{\bf nm}J_{\bf nm}\left(s^z_{\bf n}s^z_{\bf
m}+s^{+1}_{\bf n}s^{-1}_{\bf m}\right)-\mu\sum_{\bf n}X_{\bf n},
\label{hamiltonian}
\end{equation}
where $a_{\bf n\sigma}=|{\bf n}\sigma\rangle\langle{\bf n}0|$ is the
hole annihilation operator, {\bf n} and {\bf m} label sites of the
square lattice, $\sigma=\pm 1$ is the spin projection, $|{\bf
n}\sigma\rangle$ and $|{\bf n}0\rangle$ are site states corresponding
to the absence and presence of a hole on the site. For nearest neighbor
interactions $t_{\bf nm}=t\sum_{\bf a}\delta_{\bf n,m+a}$ and $J_{\bf
nm}=J\sum_{\bf a}\delta_{\bf n,m+a}$ where $t$ and $J$ are the hopping
and exchange constants and the four vectors {\bf a} connect nearest
neighbor sites. The spin-$\frac{1}{2}$ operators can be written as
$s^z_{\bf n}=\frac{1}{2}\sum_\sigma\sigma|{\bf
n}\sigma\rangle\langle{\bf n}\sigma|$ and $s^\sigma_{\bf n}=|{\bf
n}\sigma\rangle\langle{\bf n},-\sigma|$, $\mu$ is the chemical
potential and $X_{\bf n}=|{\bf n}0\rangle\langle{\bf n}0|$.

The method suggested in Ref.~7 is based on Mori's projection operator
technique\cite{Mori} which allows one to represent Green's functions in
the form of continued fractions and gives a way for calculating their
elements. The residual term of the fraction can be approximated by the
decoupling which reduces this many-particle Green's function to a
product of simpler functions. In this way we obtained the following
self-energy equations for the hole $G({\bf k}t)=
-i\theta(t)\langle\{a_{\bf k\sigma}(t),a^\dagger_{\bf k\sigma}\}
\rangle$ and spin $D({\bf k}t)=-i\theta(t)\langle[s^z_{\bf k}(t),
s^z_{\bf -k}]\rangle$ Green's functions:
\begin{eqnarray}
D({\bf k}\omega)&=&\frac{[4J\alpha(\Delta+1+\gamma_{\bf
k})]^{-1}\Pi({\bf k}\omega)+4JC_1(\gamma_{\bf k}-1)}{\omega^2-\Pi({\bf
k}\omega)- \omega^2_{\bf k}}, \nonumber\\[-1ex]
&&\label{se} \\[-1ex]
G({\bf k}\omega)&=&\phi[\omega-\varepsilon_{\bf k}+\mu'-\Sigma({\bf
k}\omega)]^{-1}, \nonumber
\end{eqnarray}
where the self-energies $\Pi({\bf k}\omega)$ and $\Sigma({\bf
k}\omega)$ read
\begin{eqnarray}
{\rm Im}\,\Pi({\bf k}\omega)&=&\frac{16\pi t^2J}{N}(\Delta+1+
 \gamma_{\bf k})\sum_{\bf k'}(\gamma_{\bf k}-\gamma_{\bf
 k+k'})^2\int^\infty_{-\infty}d\omega'[n_F(\omega+\omega')-n_F(\omega')]
 \nonumber\\
&&\times A({\bf k+k'},\omega+\omega')A({\bf k'}\omega'), \nonumber\\[-1ex]
&&\label{po}\\[-1ex]
{\rm Im}\,\Sigma({\bf k}\omega)&=&\frac{16\pi t^2}{N\phi}\sum_{\bf k'}
 \int_{-\infty}^\infty d\omega'\left[\gamma_{\bf k-k'}+\gamma_{\bf
 k}+{\rm sgn}(\omega')(\gamma_{\bf
 k-k'}-\gamma_{\bf k})\sqrt{\frac{1+\gamma_{\bf k'}}{1-\gamma_{\bf
 k'}}}\right]^2 \nonumber\\
&&\times [n_B(-\omega')+n_F(\omega-\omega')
 ]A({\bf k-k'},\omega-\omega')B({\bf k'}\omega'). \nonumber
\end{eqnarray}
In the above formulas $\gamma_{\bf k}=\frac{1}{4}\sum_{\bf a}\exp(i{\bf
ka})$, $n_F(\omega)=[\exp(\omega/T)+1]^{-1}$, $n_B(\omega)=\left[
\exp(\omega/T)-1\right]^{-1}$, $A({\bf k}\omega)=-\pi^{-1}{\rm
Im}\,G({\bf k}\omega)$ and $B({\bf k}\omega)= -\pi^{-1}{\rm Im}\,D({\bf
k}\omega)$ are the hole and spin spectral functions,
$\mu'=\mu-4F_1\phi^{-1}t+3C_1\phi^{-1}J$, $\phi=\frac{1}{2}(1+x)$ and
\begin{equation}
\omega^2_{\bf k}=16J^2\alpha|C_1|(1-\gamma_{\bf k})
(\Delta+1+\gamma_{\bf k}),\quad \varepsilon_{\bf k}=(4\phi
t+6C_1\phi^{-1}t-3F_1\phi^{-1}J)\gamma_{\bf k}, \label{seed}
\end{equation}

The hole concentration $x$, the spin $C_1=\langle s_{\bf l}^{+1}s_{\bf
l+a}^{-1}\rangle$ and hole $F_1=\langle a_{\bf l}^\dagger a_{\bf
l+a}\rangle$ nearest neighbor correlations are determined by the
relations
\begin{eqnarray}
&&x=N^{-1}\sum_{\bf k}\int_{-\infty}^\infty d\omega n_F(\omega)A({\bf
 k}\omega),\quad
C_1=\frac{2}{N}\sum_{\bf k}\gamma_{\bf k}\int_0^\infty d\omega
 \coth\!\left(\frac{\omega}{2T}\right)B({\bf k}\omega),\nonumber\\[-1ex]
&&\label{xcf}\\[-1ex]
&&F_1=N^{-1}\sum_{\bf k}\gamma_{\bf k}\int_{-\infty}^\infty d\omega
 n_F(\omega)A({\bf k}\omega).\nonumber
\end{eqnarray}
The parameter of vertex correction $\alpha$ is set equal to its value
in the undoped case, with introduced artificial broadening
$\alpha=1.802$. As known,\cite{Mermin} in the considered
two-dimensional system the long-range antiferromagnetic ordering is
destroyed at any nonzero $T$ and, as can be seen from the above
formulas, at any nonzero $x$. The arising state with short-range order
has to satisfy the constraint of zero site magnetization $\langle
s^z_{\bf l}\rangle=0$ which can be written in the form
\begin{equation}\label{zsm}
\frac{1}{2}(1-x)=\frac{2}{N}\sum_{\bf k}\int_0^\infty d\omega
\coth\!\left(\frac{\omega}{2T}\right)B({\bf k}\omega).
\end{equation}
This equation fixes the parameter $\Delta$ describing the spin gap at
$(\pi,\pi)$.

The same derivation applied to the transversal spin Green's function
$\langle\!\langle s_{\bf k}^{-1}\big|s_{\bf k}^{+1}\rangle\!\rangle$
gives $\langle\!\langle s_{\bf k}^{-1}\big|s_{\bf k}^{+1}\rangle\!
\rangle=2D({\bf k}\omega)$ indicating that the approximation used
retains properly the rotation symmetry of spin components in the
paramagnetic state.

For low $x$ and $T$ the magnitude of the dispersion $\varepsilon_{\bf
k}$ is small in comparison with $t$ manifesting the band narrowing in
the antiferromagnetic surrounding.

Equations~(\ref{se})--(\ref{zsm}) form a closed set which can be solved
by iteration. We carried out such calculations for a 20$\times$20
lattice and the parameters $t=0.5$~eV, $J=0.1$~eV corresponding to
cuprates. Small artificial broadenings $-\eta$ and $-2\eta\omega_{\bf
k}$, $\eta=0.02t$, were added to ${\rm Im}\Sigma({\bf k}\omega)$ and
${\rm Im}\Pi({\bf k}\omega)$, respectively, to widen narrow lines and
to stabilize the iteration procedure.

The frequencies of spin excitations satisfy the equation
\begin{equation}
\omega^2-{\rm Re}\Pi({\bf k}\omega)-\omega^2_{\bf k}=0 \label{sef}
\end{equation}
[see Eq.~(\ref{se})]. For low $x$ and $T$ their dispersion is close the
dispersion of spin waves (see Fig.~\ref{Fig_i}a). The main difference
is the spin gap at $(\pi,\pi)$ the magnitude of which grows with $x$
and $T$ (Fig.~\ref{Fig_i}b).
\begin{figure}
\caption{The dispersion of spin
excitations. Vertical bars show decay widths $|{\rm Im}\Pi({\bf
k}\omega)|/(2\omega_{\bf k})$. Points $Y$, $M$ and $S$ corresponds to
${\bf k}=(0,\pi)$, $(\pi,\pi)$ and $(\pi/2,\pi/2)$,
respectively.\label{Fig_i}}
\end{figure}
In an infinite crystal this gap is directly connected with the spin
correlation length $\xi$. Indeed, for large distances and low $T$ we
find
\begin{equation}\label{spincor}
\left\langle s^z_{\bf l}s^z_{\bf 0}\right\rangle=N^{-1}\sum_{\bf k}{\rm
e}^{i\bf kl}\int_0^\infty d\omega\coth\left(\frac{\omega}{2T}\right)
B({\bf k}\omega)\:\propto\:{\rm e}^{i\bf Ql}(\xi/|{\bf l}|)^{1/2}{\rm
e}^{-|{\bf l}|/\xi},
\end{equation}
where ${\bf Q}=(\pi,\pi)$ and $\xi=a/(2\sqrt{\Delta})$ with the
intersite distance $a$. For low $x$ we found that $\Delta\approx 0.2x$
and consequently $\xi\approx a/\sqrt{\Delta}$. This relation has been
experimentally observed\cite{Keimer} in La$_{2-x}$Sr$_x$CuO$_4$.

As seen from Fig.~\ref{Fig_i}b, with growing $x$ the spin excitation
branch is destroyed in some region around the $\Gamma$ point --
Eq.~(\ref{sef}) has no solution for real $\omega$ due to negative ${\rm
Re}\Pi({\bf k}\omega)$. For fixed $T$ the size of this region grows
with $x$. However, as follows from Fig.~\ref{Fig_i}b, with increasing
$T$ the branch is recovered in this region. This result is the
consequence of the temperature broadening of the spin-polaron peak in
$A({\bf k}\omega)$.\cite{Sherman98,Sherman02} Large negative values of
${\rm Re}\Pi({\bf k}\omega)$ which are the reason for the lack of a
solution of Eq.~(\ref{sef}) are connected with this peak. With its
broadening $|{\rm Re}\Pi|$ becomes smaller.

The spin correlations $C_{ml}=\left\langle s^z_{\bf l}s^z_{\bf 0}\right
\rangle$, ${\bf l}=(n,m)$, Eq.~(\ref{spincor}), are shown in
Fig.~\ref{Fig_ii}.
\begin{figure}
\caption{(a) Spin correlations vs. $x$ for
$T=0.02t$. (b) Spin correlations along the diagonal [i.e., ${\bf
l}=(m,m)$] of the crystal for $x=0.12$. The respective temperatures are
indicated near the curves.\label{Fig_ii}}
\end{figure}
Distant correlations decrease rapidly with $x$. For large enough $x$
and $T$ the correlations decay exponentially with distance in the
considered finite lattice.

The magnetic susceptibility is connected with the spin Green's function
(\ref{se}) by the relation $\chi^z({\bf k}\omega)=-4\mu_B^2 D({\bf
k}\omega)$, where $\mu_B$ is the Bohr magneton. Experiments on
inelastic neutron scattering give information on the dynamic spin
susceptibility which can be directly compared with the calculated
results. Such comparison is carried out in Fig.~\ref{Fig_iii}.
\begin{figure}
\caption{The imaginary
part of the spin susceptibility for ${\bf k}=(\pi,\pi)$. Curves
demonstrate results of our calculations for $T=0.02t$, and $x=0.043$,
0.08 and 0.12 (from top to bottom). Squares show the imaginary part of
the odd spin susceptibility for ${\bf k}=(\pi,\pi)$
measured\protect\cite{Fong} in normal state YBa$_2$Cu$_3$O$_{7-y}$ at
$T=100$~K for $y=0.5$, 0.17 and 0.03 (from top to bottom).
\label{Fig_iii}}
\end{figure}
YBa$_2$Cu$_3$O$_{7-y}$ is a bilayer crystal and the symmetry allows one
to divide the susceptibility into odd and even parts. For the
antiferromagnetic intrabilayer coupling the odd part of the
susceptibility can be compared with the calculated results. The oxygen
deficiencies $y=0.5$, 0.17 and 0.03 in the experimental data in
Fig.~\ref{Fig_iii} correspond to the hole concentrations $x=0.07$, 0.13
and 0.17 respectively.\cite{Tallon} We scale the experimental ${\rm
Im}\chi$ so that their maximal values coincide approximately with the
calculated maxima. In absolute units the calculated maximal values are
approximately 2--2.5 times larger than the experimental ones. As seen
from Fig.~\ref{Fig_iii}, the calculated data reproduce correctly the
frequency dependence of the susceptibility, the values of the frequency
for which ${\rm Im}\chi(\pi,\pi)$ reaches maximum and their evolution
with doping. We notice also that the calculated temperature variation
of the susceptibility is in good agreement with experiment.

The calculated dependencies of the uniform static spin susceptibility
$\chi_0=\chi({\bf k}\rightarrow 0,\omega=0)$ on $T$ and $x$ are shown
in Fig.~\ref{Fig_iv}.
\begin{figure}
\caption{\label{Fig_iv} The uniform static spin
susceptibility vs.\ $T$ (a) and $x$ (b).}
\end{figure}
In this figure $\chi_0/\mu_B^2$ lies in the range 2-2.6~eV$^{-1}$ which
is close to the values 1.9--2.6~eV$^{-1}$ obtained\cite{Barzykin} for
YBa$_2$Cu$_3$O$_{7-y}$. The temperature dependence of $\chi_0$ has a
maximum and its temperature $T_m$ grows with decreasing $x$. Analogous
behavior of $\chi_0$ was observed\cite{Johnston,Takigawa} in cuprates
for large enough $x$. In Fig.~\ref{Fig_iv}a $T_m\approx 600$~K which is
close to $T_m$ observed\cite{Johnston} in La$_{2-x}$Sr$_x$CuO$_4$ for
comparable $x$. As known, in the undoped antiferromagnet $T_m\approx
J$. On the high-temperature side from the maximum $\chi_0(T)$ tends to
the Curie-Weiss dependence $1/T$. As seen from Fig.~\ref{Fig_iv}a, the
two curves for the different $x$ are very close in shape and can be
superposed by scaling to the same maximal value of $\chi_0$ and $T_m$.
Analogous scaling was observed\cite{Johnston} in
La$_{2-x}$Sr$_x$CuO$_4$. As in experiment,\cite{Johnston} the
dependence $\chi_0(x)$ in Fig.~\ref{Fig_iv}b has a maximum. However,
the experimental maximum is achieved at $x\approx 0.25$.

It is of interest to clarify the reason for the nonmonotonic behavior
of the dependence $\chi_0(T)$. From Eqs.~(\ref{se}) and (\ref{po}) it
is clear that $\chi({\bf k},\omega=0)$ is continuous for small {\bf k}.
Therefore let us consider a small but finite value of {\bf k} for which
$\chi({\bf k},0)\propto\int_{-\infty}^\infty d\omega' B({\bf
k}\omega')/\omega'$. The maximum in $B({\bf k}\omega')$ is shifted to
lower frequencies and its intensity decreases with increasing
temperature. However this dependence is superimposed with the function
$1/\omega'$ in the integral. Nonmonotonic behavior of $\chi_0(T)$ is
the consequence of the existence of some optimal location of the
maximum in $B({\bf k}\omega')$ when the area under the curve $B({\bf
k}\omega')/\omega'$ is maximal. Johnston's scaling\cite{Johnston} means
that the temperature softening of the long-wavelength maximum occurs
similarly for different $x$ and that holes and temperature fluctuations
act in a similar manner as sources of the softening.

The spin-lattice relaxation and spin-echo decay rates were calculated
with the use of the equations\cite{Barzykin}
\begin{eqnarray}
&&\frac{1}{^\alpha T_{1\beta}T}=\frac{1}{2\mu^2_BN}\sum_{\bf k}
 \,^\alpha\!
 F_\beta({\bf k})\frac{{\rm Im}\,\chi({\bf k}\omega)}{\omega},
 \quad\omega\rightarrow 0,\nonumber\\[-1ex]
&&\label{nmr}\\[-1ex]
&&\frac{1}{^{63}T_{2G}^2}=\frac{0.69}{128\mu_B^4}\left\{
 \frac{1}{N}\sum_{\bf k}\,^{63}\!F_e^2({\bf k})\left[{\rm Re}\,
 \chi({\bf k}0)
 \right]^2-\left[\frac{1}{N}\sum_{\bf k}\,^{63}\!F_e({\bf k})
 {\rm Re}\,\chi({\bf k}0)\right]^2\right\}, \nonumber
\end{eqnarray}
where the form factors $^\alpha\!F_\beta({\bf k})$ and hyperfine
coupling constants were taken from Ref.~4. The superscript $\alpha=63$
or 17 indicates that the respective quantity belongs to Cu and O,
respectively. The subscripts $\beta=\|$ or $\bot$ refer to the
direction of the applied static magnetic field {\bf H} with respect to
the axis {\bf c} perpendicular to the Cu-O plane. The form factor
$^{63}\!F_e$ is the filter for the Cu spin-echo decay time
$^{63}T_{2G}$. Our calculated results and the respective experimental
data are given in Fig.~\ref{Fig_v}.
\begin{figure}
\caption{The temperature
dependencies of the spin-lattice relaxation and spin-echo decay rates.
Open circles with right axes represent experimental results, filled
circles with left axes are our calculations. (a,c,d) calculations for
${\bf H\|c}$ and $x=0.12$, measurements\protect\cite{Takigawa} in
YBa$_2$Cu$_3$O$_{6.63}$ ($x\approx 0.1$).\protect\cite{Tallon} (b)
calculations for nonoriented configuration with $x=0.043$,
measurements\protect\cite{Imai} in La$_{1.96}$Sr$_{0.04}$CuO$_4$.
\label{Fig_v}}
\end{figure}
The calculations reproduce satisfactorily main peculiarities of the
temperature dependence of the spin-lattice and spin-echo decay rates.
The growth of $(^{63}T_1T)^{-1}$ with decreasing $x$ is connected with
the increase of the spectral intensity of spin excitations near
$(\pi,\pi)$. For the same $x$ $(^{63}T_1T)^{-1}$ is one to two orders
of magnitude larger than $(^{17}T_1T)^{-1}$. This is a consequence of
the spin susceptibility which is strongly peaked near $(\pi,\pi)$ and
the form factors which test different {\bf k} regions.\cite{Barzykin}

In Fig.~\ref{Fig_v} the calculated values of the spin-lattice
relaxation rates are nearly an order of magnitude smaller than the
experimental ones. This difference is connected with the approximation
made in the calculation of $D({\bf k}\omega)$ the continued fraction of
which was cut short in the second link. This leads to an additional
power of $\omega$ in the denominator of $B({\bf k}\omega)$.

It is of interest to clarify the reason for the nonmonotonic
temperature dependence of $(^{63}T_1T)^{-1}$ for moderate $x$ (see
Fig.~\ref{Fig_v}a). The main contribution to $(^{63}T_1T)^{-1}$ is
given by the vicinity of the $(\pi,\pi)$ point. For low $\omega$ in
this region ${\rm Im}\chi({\bf k}\omega)$ first grows then decreases
with increasing $T$. This is connected with two competing tendencies in
the spin spectral function: i) the spin gap $\Delta$ grows with $T$
which leads to a decrease of $B$ in the mentioned regions of {\bf k}
and $\omega$ and ii) the temperature broadening of the maximum in $B$
which leads to its growth in these regions. For moderate $x$ and low
$T$ practically $\Delta$ does not depend on $T$, the second tendency
dominates and $(^{63}T_1T)^{-1}$ grows with $T$. For higher $T$ the gap
grows more rapidly with $T$ leading to the predominance of the first
tendency and $(^{63}T_1T)^{-1}$ decreases with $T$. For small $x$
temperature fluctuations dominate, $\Delta$ grows noticeably even for
small $T$ and $(^{63}T_1T)^{-1}$ decreases monotonously with $T$ (see
Fig.~\ref{Fig_v}b). Due to the form factor $^{17}\!F_\|({\bf k})$, the
central part of the Brillouin zone makes the main contribution to
$(^{17}T_1T)^{-1}$. For moderate $T$ and $x$ the existence of the spin
gap affects $B({\bf k}\omega)$ weakly here. For small $\omega$ the
temperature dependence of $B({\bf k}\omega)$ is determined by
broadening. As a consequence, $(^{17}T_1T)^{-1}$ grows monotonously
with $T$ (see Fig.~\ref{Fig_v}c). The low-temperature decreases of
$(^{63}T_1T)^{-1}$ and $\chi^0$ were considered as manifestations of
the spin gap. As follows from the above discussion, these peculiarities
are not directly connected with the gap.

This work was partially supported by the ESF grant No.~4022 and by DFG
(SFB 393).


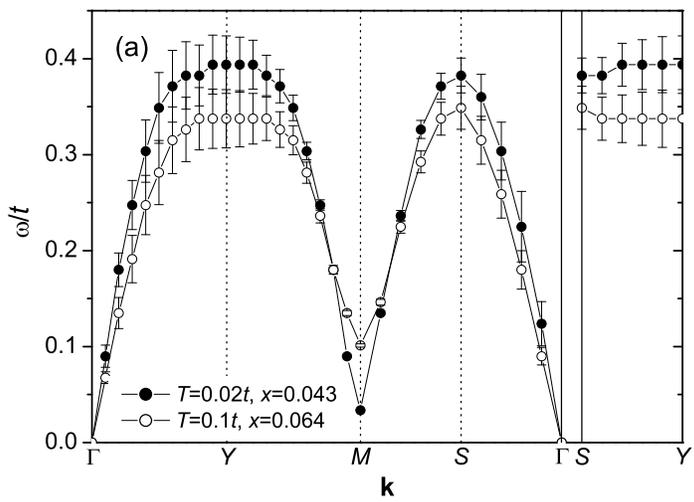
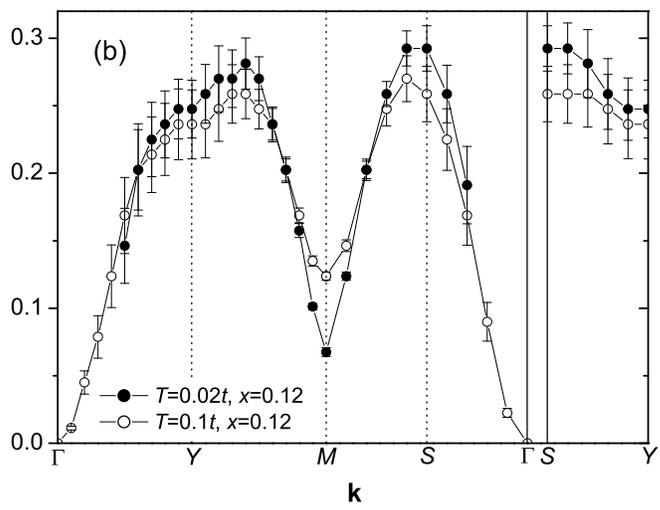
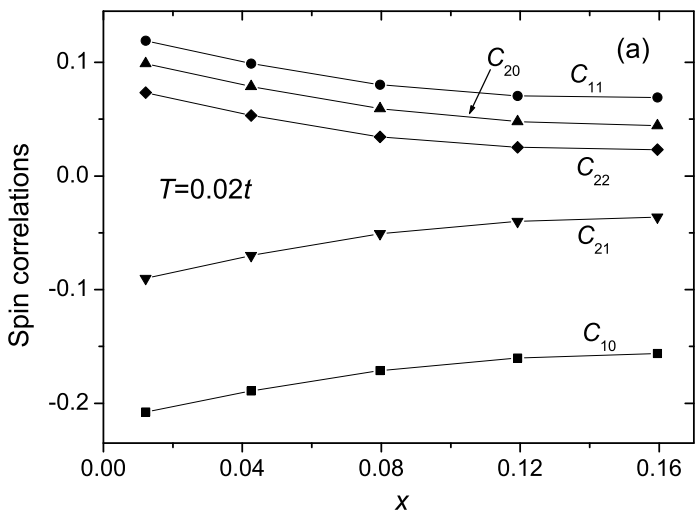
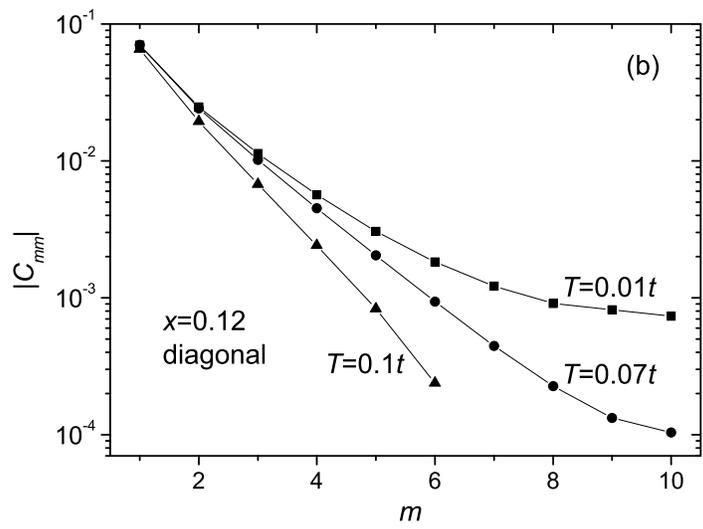
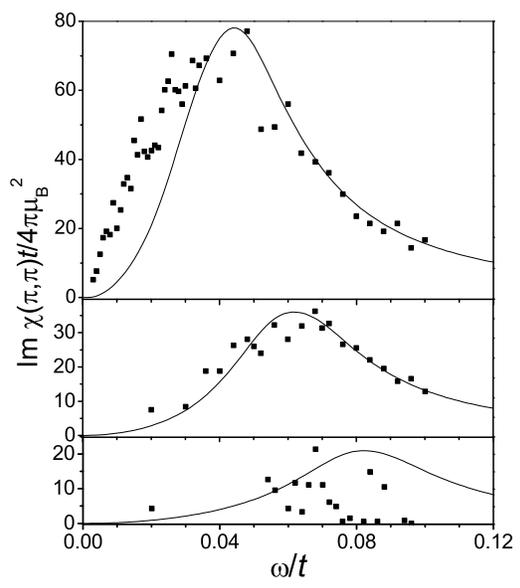
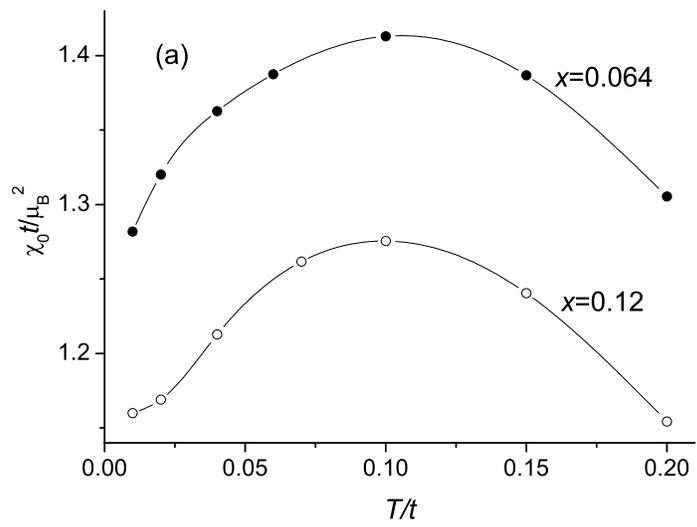
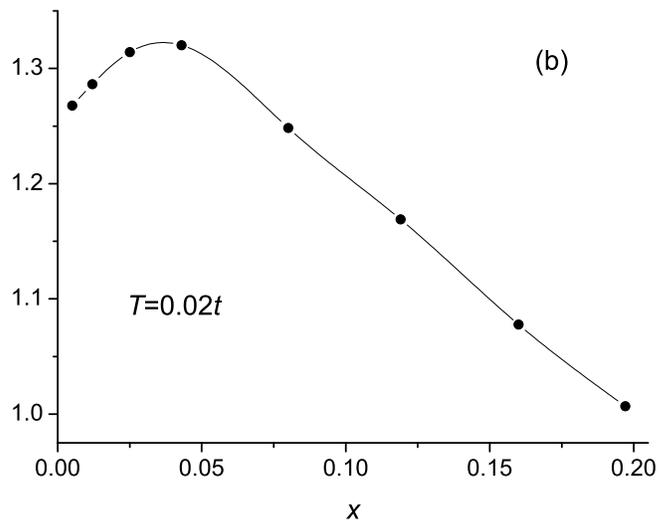
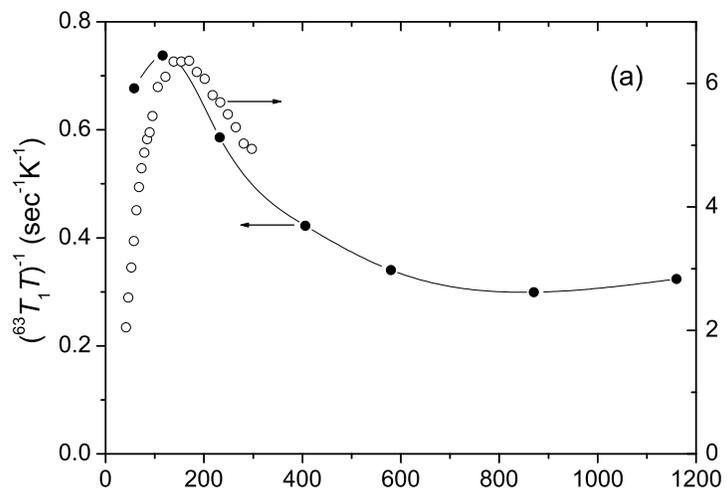
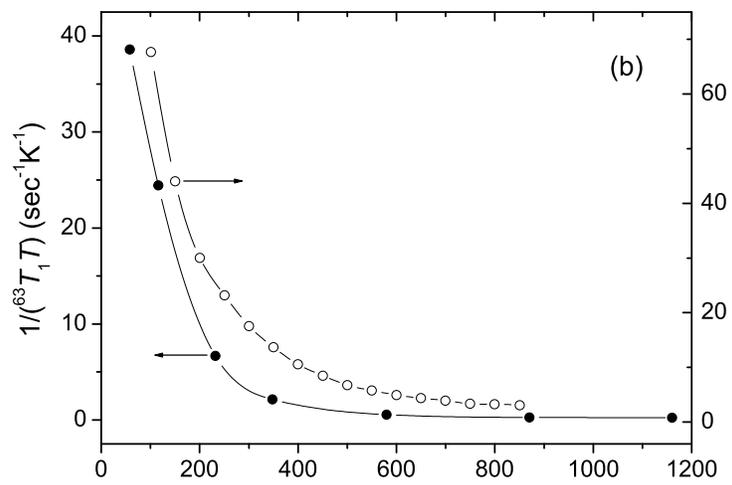
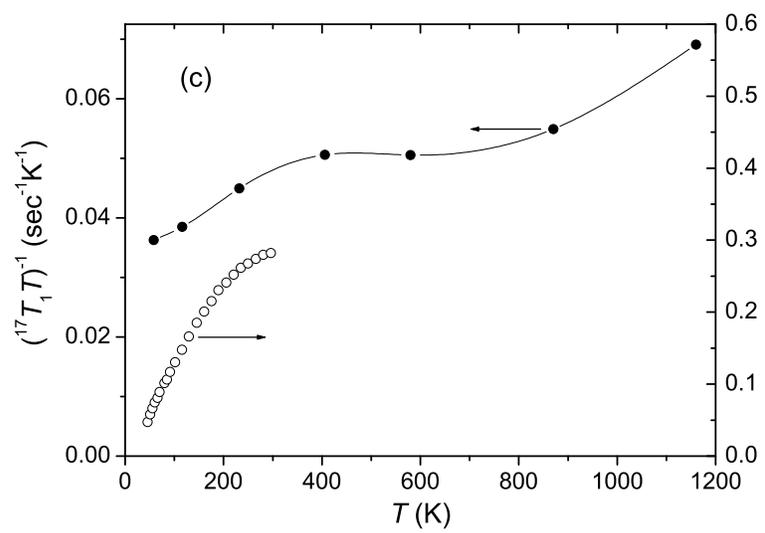
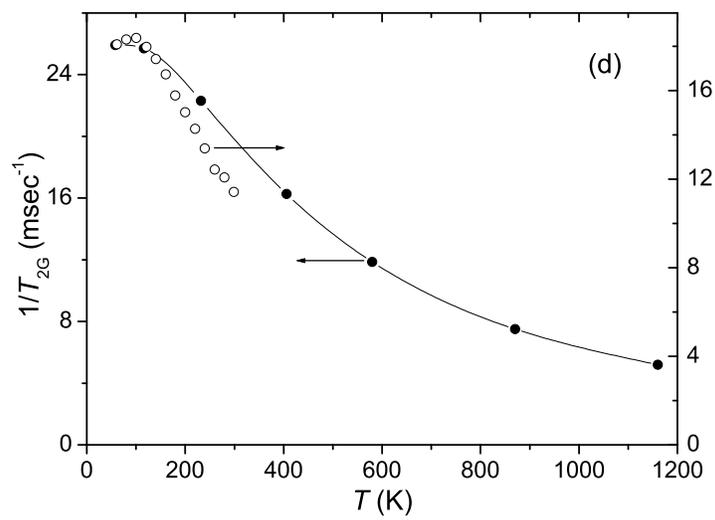
\end{document}